\documentclass[12pt]{iopart}

\usepackage[german,english]{babel}
\usepackage{amssymb}
\usepackage{amsfonts}
\usepackage{mathrsfs}
\usepackage{epsfig}

\newcommand {\be} {\begin{equation}}   
\newcommand {\ee} {\end{equation}}
\newcommand {\bea} {\begin{eqnarray}}
\newcommand {\eea} {\end{eqnarray}}
\newcommand {\bes} {\begin{displaymath}}
\newcommand {\ees} {\end{displaymath}}

\begin{document}

\title[Directed motion from coupled random processes]{Directed motion emerging
from two coupled random processes: Translocation of a chain through a
membrane nanopore driven by binding proteins}

\author{Tobias Ambj\"ornsson, Michael A. Lomholt and Ralf Metzler}
\address{NORDITA -- Nordic Institute for Theoretical Physics,\\ 
Blegdamsvej 17, DK-2100 Copenhagen \O, Denmark}
\ead{\mailto{ambjorn@nordita.dk}, \mailto{mlomholt@nordita.dk}, \mailto{metz@nordita.dk}}

\begin{abstract}
We investigate the translocation of a stiff polymer consisting of $M$
monomers through a nanopore in a membrane, in the presence of binding
particles (chaperones) that bind onto the polymer, and partially
prevent backsliding of the polymer through the pore. The process is
characterized by the rates: $k$ for the polymer to make a diffusive
jump through the pore, $q$ for unbinding of a chaperone, and the rate
$q\kappa$ for binding (with a binding strength $\kappa$); except for
the case of no binding $\kappa=0$ the presence of the chaperones give
rise to an effective force that drives the translocation process. In
more detail, we develop a dynamical description of the process in
terms of a (2+1) variable master equation for the probability of
having $m$ monomers on the target side of the membrane with $n$ bound
chaperones at time $t$. Emphasis is put on the calculation of the mean
first passage time $\intercal$ as a function of total chain length
$M$. The transfer coefficients in the master equation are determined
through detailed balance, and depend on the relative chaperone size
$\lambda$, binding strength $\kappa$, as well as the two rate
constants $k$ and $q$. The ratio $\gamma=q/k$ between the two rates
determines, together with $\kappa$ and $\lambda$, three limiting
cases, for which analytic results are derived: (i) For the case of
slow binding ($\gamma\kappa\rightarrow 0$), the motion is purely
diffusive, and $\intercal\simeq M^2$ for large $M$; (ii) for fast
binding ($\gamma\kappa\rightarrow \infty$) but slow unbinding ($\gamma
\rightarrow 0$), the motion is, for small chaperones $\lambda=1$,
ratchet-like, and $\intercal\simeq M$; (iii) for the case of fast
binding and unbinding dynamics ($\gamma \rightarrow \infty$ and $
\gamma\kappa\rightarrow \infty$), we perform the adiabatic elimination
of the fast variable $n$, and find that for a very long polymer
$\intercal\simeq M$, but with a smaller prefactor than for
ratchet-like dynamics. We solve the general case numerically as a
function of the dimensionless parameters $\lambda$, $\kappa$ and
$\gamma$, and compare to the three limiting cases.
\end{abstract}


\maketitle

\section{Introduction}

The passage of a biopolymer (DNA, RNA, proteins) through a nanopore
embedded in a membrane is one of the most crucial processes in
biology, examples including: the translocation of proteins through the
endoplasmatic reticulum, the passage of RNA through the nucleus pore
membrane, the viral injection of DNA into a host, as well as DNA
plasmid transport from cell to cell through cell walls
\cite{alberts}. Moreover, biotechnology applications of translocation
processes such as rapid DNA sequencing \cite{kasia,meller}, secondary
structure determination of RNA \cite{gerland}, as well as analyte
detection and nanosensing \cite{kasia1,ctn,dimarzio} have been
discussed.  Goals such as drug delivery are among the ultimate
research directions connected to translocation.

Experimentally, biopolymer translocation can be probed on the single
molecule level. In such experiments one commonly employs an
$\alpha$-hemolysin protein pore introduced into a planar lipid
bilayer. But, of late, nanopores can also be created in a well-defined
manner by etching techniques in silicon oxide \cite{dekker}. In vitro,
translocation experiments involve either a constant driving voltage
maintained during the entire measurement \cite{meller,kasia}; or a
high applied voltage for threading the head of the biopolymer through
the pore, followed by an off-voltage and a low probe voltage
\cite{bates}. Once a biopolymer enters the pore, it blocks the
electrical current of ions through the pore, as monitored by
patch-clamp techniques \cite{mellerrev}. From such measurements one
may directly extract, for instance, the mean translocation time
$\intercal$ and its dependence on applied voltage and on other
physical parameters of the system. Note that cross-membrane potentials
driving translocation also exist in nature \cite{alberts}.

\begin{figure}
\begin{center}
\scalebox{0.6}{\epsfbox{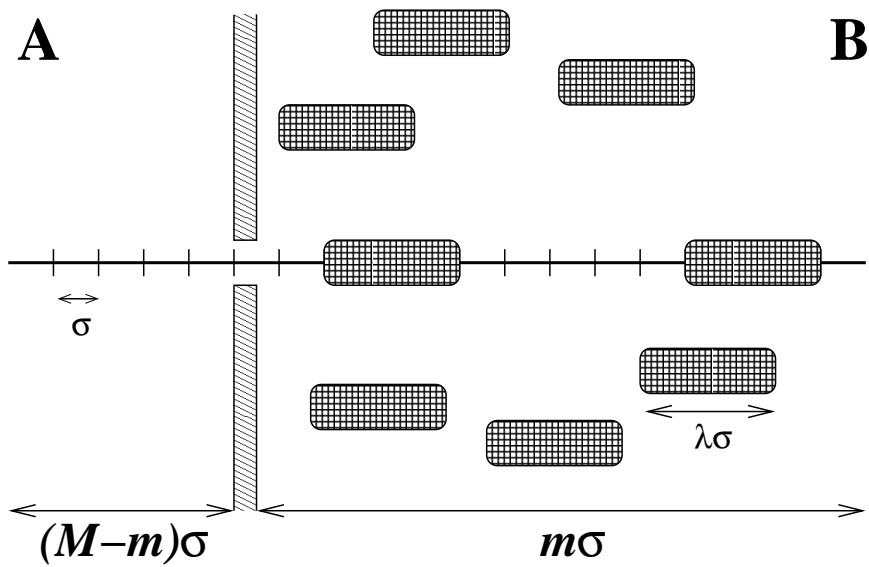}}
\end{center}
\caption{Schematic of the translocation process: Binding proteins
  (chaperones), of volume concentration $c_0$, attempt to bind to the
  translocating chain. Once bound, chaperones (partially) rectify the
  passage of the biopolymer through the nanopore. The strength of the
  binding between the polymer and the chaperones is characterized by
  the dimensionless parameter $\kappa=c_0 K^{\rm eq}$, where $K^{\rm
    eq}$ is the chaperone binding constant, see text. When bound onto
  the polymer a single chaperone occupies a length $\lambda\sigma$,
  where $\sigma$ is the linear monomer size (size of a base,
  base-pair, or aminoacid).  The chaperones span several monomeric
  unit lengths, typically $\lambda\sim 10-20$. In the absence of
  chaperones the polymer diffuses through the pore with a diffusive
  rate $k$ (the bare diffusion constant for the chain is thus
  $D_0=\sigma^2 k$). The (un)binding process is characterized by an
  unbinding rate $q$ for a single chaperone, leading to a competition
  of two random processes depending on the ratio $\gamma=q/k$.}
\label{fig:scheme}
\end{figure}

Another in vivo mechanism is the driving by binding proteins,
so-called chaperones, that appears to be particularly common for
protein translocation
\cite{zandi,liebermeister,elston,simon,rapoport,neupert}, but is also
implicated in the DNA transport through membranes
\cite{salman,farkas}. The driving of the translocation process by
binding proteins constitutes a fine example of how directed motion
can emerge from the coupling of two random processes. Schematically,
this is illustrated in figure \ref{fig:scheme}: binding
proteins on the trans side (B) of the membrane bind nonspecifically to
the already translocated portion of the biopolymer. One expects three
limiting dynamical regimes: (i) for slow binding (in
reference \cite{ambme} referred to as the {\it diffusive regime}), the
relative diffusive motion of the polymer through the pore is so fast
that the chaperones do not have time to bind; (ii) for fast binding
but slow unbinding (the {\em irreversible binding regime}), the motion
is ratchet-like: The chaperones bind (infinitely) fast and while being
attached to the polymer, they prevent backsliding through the pore,
and therefore (partially) rectify the diffusive process. (iii)
for the case of fast binding and unbinding dynamics (the {\em
  reversible binding regime}), one may assume that the (fast)
chaperones exert an effective force originating from the chaperone
chemical potential difference between the trans and cis sides
\cite{sung,lubensky,muthu,muthu1,kafri}. Given the rather short
biopolymer segments used in typical experimental setups \cite{meller}
as well as short translocating biopolymers in vivo \cite{alberts}, in
combination with the rather large binding interface between chaperones
and driven biopolymer, non-negligible finite size effects occur, that
require a modeling beyond the effective chemical potential. Such
finite size effects were studied in detail in reference \cite{ambme},
finding distinct oscillations in the effective driving force. Although
the different dynamical regimes above have been investigated in
detail, so far no model has been able to account for all regimes and
the switching in between them, as tuned by changing the physical
parameters, that is possible in in vitro experiments. A first step
towards such a unification was taken in reference \cite{zandi} where
Brownian dynamics simulation were performed, and compared to the
results obtained from a (2+1)-variable Fokker-Planck equation.

We here extend the work in reference \cite{zandi} and propose a
(2+1)-variable master equation formalism for describing the full
coupled dynamics of chaperone (un)binding and polymer diffusion during
the translocation. In contrast to the continuum description in
reference \cite{zandi}, our discrete approach can explicitly account
for the ratchet mechanism, and include chaperones which are larger
than the size of a monomer. Our approach resembles recent models
developed for the coupling of the dynamics of DNA denaturation zones
and selectively single-stranded DNA binding proteins
\cite{ssb,ssb1}. This general scheme allows us, as limiting cases, to
consider the different regimes of fast and slow chaperone (un)binding
mentioned above, including the explicit limit case of pure ratcheting.
In that course, we concentrate on the effect of the chaperones, and we
view the biopolymer as a stiff chain.  Additional entropic effects due
to the accessible degrees of freedom of the chain can be
straightforwardly incorporated, as is briefly discussed. We also do
not consider anomalous diffusion dynamics of very long translocating
chains here \cite{yacov,yacov1,translo,report}. We note that the
master equation approach is a natural powerful approach to
translocation dynamics (and similar random processes in systems with a
discrete coordinate), as was demonstrated in references
\cite{flomenbom,flomenbom1}.

We finally point out three differences between our approach compared to
the original Brownian ratchet ideas put forward in reference
\cite{simon,peskin}. (1) In \cite{simon,peskin} the polymer motion
through the pore was assumed to proceed through continuous diffusive
motion until a chaperone binds; an implicit assumption in that
approach is that the characteristic diffusive step length is much
smaller than the size of a chaperone. In contrast, we assume that the
motion proceeds through random jumps of length $\sigma$ with a rate
$k$, where $\sigma$ is taken to be of the order the length of a base,
base-pair or aminoacid. The two approaches are identical when all
other length scales of the problem are much larger than $\sigma$;
however in our problem the size of the chaperones can be of the order
of $\sigma$, which in the general case (and in particular for the case
of strong binding) alters the dynamics. The justification for our
approach lies in that experimentally it is found that the mean
translocation time of single-stranded DNA through an
$\alpha$-hemolysin pore is roughly 100 fold smaller than expected
including only hydrodynamic friction in the pore
\cite{lubensky}. Also, for protein translocation the translocation
times are many orders of  magnitude larger than expected from simple
hydrodynamic considerations \cite{chauwin}. These results suggest that
there are strong interactions between the pore and the translocating
polymer; adopting a similar explanatory description as in
\cite{lubensky} in which the translocation process proceeds through
jumps in a saw-tooth-like pore potential, our rate constant $k$ is
simply the Kramer's rate associated with a jump in this potential and
$\sigma$ is the distances between minima. (2) In reference
\cite{simon,peskin} the chaperones bind independently along the
polymer. Here, we include the fact that typically chaperones are larger
than the size of a monomer, giving rise to a ``car parking effect''
\cite{ambme} - a chaperone can only bind if the space between already
bound chaperones equals or is larger than the size of the chaperones;
we thus extend the model in \cite{simon,peskin} in order to include
the fact that in general chaperones do not bind independently. (3) We
here consider arbitrary ratios of the relevant (un)binding rate and
the diffusive rate $k$ - this contrasts the results in references
\cite{simon,peskin} where only the cases of Brownian ratchet motion and fast
(un)binding dynamics were considered. We will point out similarities
and differences between our approach and that in references
\cite{simon,peskin} as they arise throughout the text.
 
\section{Master equation for chaperone-driven translocation}

Let us consider the general translocation dynamics for a chain of
length $M\sigma$, where $\sigma$ is the typical linear monomer (base,
base-pair, aminoacid) size. The coordinates we choose for the
description of the chaperone-driven biopolymer passage are the number
$m$ of monomers of the chain on the trans side B (the number of
monomers on the cis side A is then $M-m$) and the number $n$ of bound
chaperones, see figure \ref{fig:scheme}. The size of the binding
interface of a chaperone bound to the biopolymer is
$\lambda\sigma$. No chaperones are assumed to be present on side
A. Due to inherent thermal fluctuations the variables $m$ and $n$ are
stochastic variables and the aim is to understand the temporal
behaviour of these variables; below we use a master equation
formulation for the associated probability distribution. This
description rests on the assumption that all other variables in the
system are fast in comparison to the rate of change of the coordinates
$m$ and $n$. From the solution of this master equation any
experimental observable can be calculated. Here, the main quantity of
interest is the mean first passage time $\intercal(M)$ as a function
of the chain length $M$ and chaperone size $\lambda$, as well as of
two dimensionless parameters introduced in section
\ref{sec:partition_function}: the binding strength $\kappa$ and the
ratio $\gamma$ between the rate $k$ associated with making a diffusive
step and the rate $q$ for chaperone unbinding.  In our description,
the size of the pore and the number of monomers present in the
pore at a given time enters only through the effective rate $k$ (see
references \cite{ambjorn,lubensky,flomenbom,flomenbom1,kolmeis} for
more detailed investigations of the effects of finite-sized pores).

\begin{figure}
\begin{center}
\scalebox{0.5}{\epsfbox{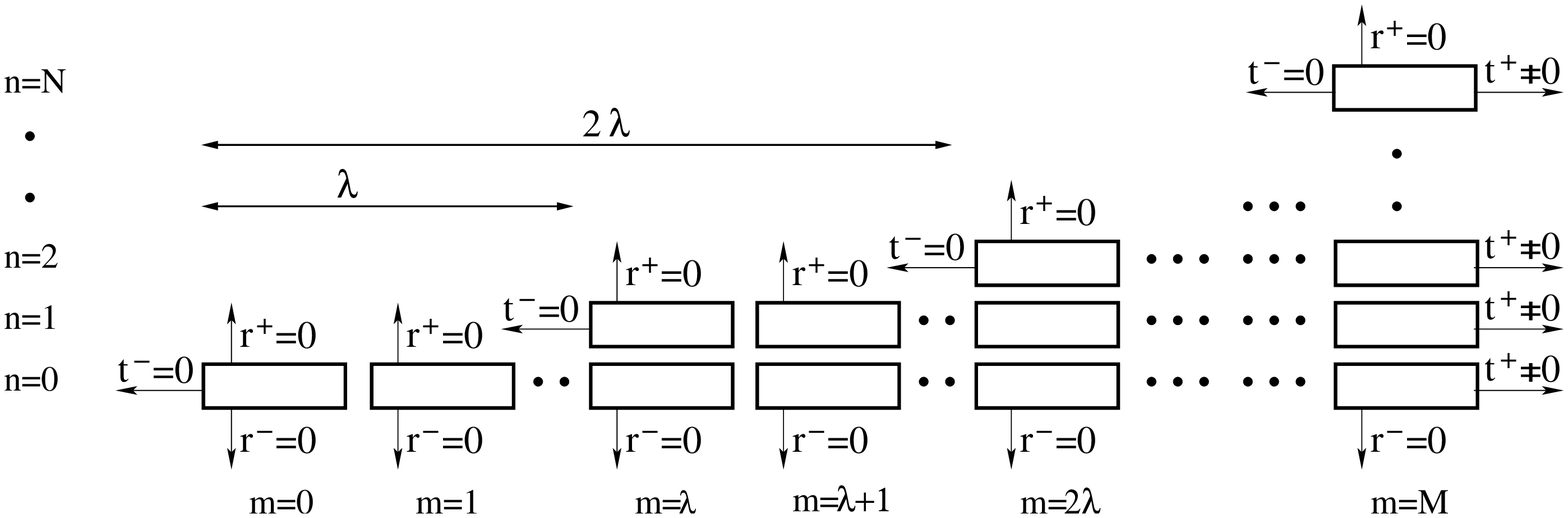}}
\end{center}
\caption{Configuration lattice spanned by translocation coordinate $m$ and number
$n$ of bound chaperones, with allowed configuration states of the system
($\Box$). Each time $m$ reaches a multiple of $\lambda$, an additional
chaperone can bind. The scheme also indicates forbidden jumps.}
\label{lattice}
\end{figure}

Denote by $P(m,n,t|m',n')$ the conditional probability density of
finding $m$ monomers of the chain on side B (the already translocated
distance in units of monomers) with $n$ bound chaperones at time $t$,
provided that at the initial time the system was in a state with
$m=m'$ and $n=n'$. With the short-hand notation
$P(m,n,t)=P(m,n,t|m',n')$, the time evolution of $P(m,n,t)$ is
governed by the (2+1)-variable master equation \cite{vankampen}
\begin{eqnarray}
\nonumber
\frac{\partial P(m,n,t)}{\partial t}&=&\mathsf{t}^+(m-1,n)P(m-1,n,t)+
\mathsf{t}^-(m+1,n,t)P(m+1,n,t)\\
\nonumber
&&\hspace{0.8cm}-\left[\mathsf{t}^+(m,n)+\mathsf{t}^-(m,n)\right]P(m,n,t)\\
\nonumber
&&+\mathsf{r}^+(m,n-1)P(m,n-1)+\mathsf{r}^-(m,n+1)P(m,n+1,t)\\
&&\hspace{0.8cm}-\left[\mathsf{r}^+(m,n)+\mathsf{r}^-(m,n)\right]P(m,n,t).
\label{master}
\end{eqnarray}
The transfer coefficients $\mathsf{t}^{\pm}$ represent the rates for a
change in the translocation coordinate $m$ and the coefficients
$\mathsf{r}^ {\pm}$ specify the (un)binding of chaperones, i.e., a
change in $n$; both transfer coefficients are explicitly defined in
section \ref{sec:partition_function}.
\footnote{For equation \eref{master} to be completely specified we
need to describe the transfer coefficients just outside the
configuration lattice (illustrated in figure \ref{lattice}) as well:
\[
\mathsf{t}^-(m=M+1,n)=\mathsf{t}^+(m=j\lambda-1,n=n^{\rm max}(m+1))=0
  \]
for integer $j$, and
  \[
\mathsf{r}^+(m,n=-1)=\mathsf{r}^-(m,n=n^{\rm max}(m)+1)=0.
  \]
}
Equation \eref{master} is subject to the general initial condition
$P(m,n,t=0|m',n')=\delta_{m,m'}\delta_{n,n'}$, where
$\delta_{z_1,z_2}$ denotes the Kronecker delta. Here, we assume that
$m'=0$ and $n'=0$:
\begin{equation}
\label{initial}
P(m,n,0)=\delta_{m,0}\delta_{n,0},
\end{equation}
i.e., that initially, the chain is fully on side A, with its head just
in the pore (and therefore has no bound chaperones), corresponding to
the typical situation in vitro and in vivo.

We now consider the boundary conditions that must be imposed on the
transfer coefficients: Once the chain has arrived completely in side
B, i.e., $m=M$, we assume that it cannot find its way back into the
pore. We impose, that is, the absorbing boundary condition
\footnote{We note that a reflecting boundary condition at $m=M$, as
  occurring for instance in the description of DNA-breathing dynamics
  \cite{ssb,ssb1}, would have the structure $\mathsf{t}^+(M,n)=0$.}
\begin{equation}
\mathsf{t}^-(M+1,n)=0.
\label{absorb}
\end{equation}
For the absorbing condition (\ref{absorb}) imposed here, the $M$th
transfer coefficient $\mathsf{t}^+(M,n)$ in general is finite, giving
rise to jumps out from the configuration lattice (see Figure
\ref{lattice}), such that in this case $\sum_{m,n}P(m,n,t)$ decreases
with time, and therefore $P(m,n,t)$ is an improper probability
distribution. At $m=0$, we impose the reflecting boundary condition
\begin{equation}
\label{reflect}
\mathsf{t}^-(0,0)=0.\label{eq:bc1}
\end{equation}
This condition guarantees that the chain does not retract from the
pore to side A. Additionally, we require that the chain cannot move
towards side A, when it is fully occupied with chaperones:
\begin{equation}
\mathsf{t}^-\left(m=n^{\mathrm{max}}(m)\lambda,n^{\mathrm{max}}(m)\right)=0,\label{eq:bc2}
\end{equation}
where 
  \be
n^{\rm max}(m)=[m/\lambda]
  \ee
is the maximum number of bound chaperones for a given $m$ (and $[
  \cdot ]$ is the Landau bracket returning the integer value of the
argument). The boundary condition \eref{eq:bc2} is crucial for the
emergence of ratchet motion as discussion in subsection
\ref{ratchet}. For the (un)binding rates, we have the natural
condition
\begin{equation}
\mathsf{r}^-(m,0)=0,\label{eq:bc3}
\end{equation}
stating that no further unbinding can occur once the chain is empty of
chaperones. Similarly, chaperones cannot bind to a fully occupied chain:
\begin{equation}
\mathsf{r}^+\left(m,n^{\mathrm{max}}(m)\right)=0.\label{eq:bc4}
\end{equation}
 The configuration lattice on which the jump dynamics, as defined
by equation \eref{master} and the transfer coefficients $\mathsf{t}^{\pm}$
and $\mathsf{r}^{\pm}$, is schematically shown in figure
\ref{lattice}, together with the boundary conditions above.

For internal points on the lattice in figure \ref{lattice} the
transfer coefficients are constrained (but not fully determined) by the
detailed balance condition \cite{vankampen}. For our specific system,
the detailed balance condition takes on the structure (compare to
references \cite{ssb,ssb1})
\begin{equation}
\label{dbt}
\mathsf{t}^+(m-1,n)\mathscr{Z}(m-1,n)=\mathsf{t}^-(m,n)\mathscr{Z}(m,n)
\end{equation}
and
\begin{equation}
\label{dbr}
\mathsf{r}^+(m,n-1)\mathscr{Z}(m,n-1)=\mathsf{r}^-(m,n)\mathscr{Z}(m,n).
\end{equation}
where $\mathscr{Z}(m,n)$ is the partition coefficient for a given $m$
and $n$. Whereas equation (\ref{dbr}) holds for all $n$, equation (\ref{dbt}) only holds for $0\le m\le M$, but not
for $m=M+1$. This is due to the fact that $\mathsf{t}^+(M,n)\neq 0$
corresponding to our absorbing boundary condition; in the case of a
reflecting boundary condition, $\mathsf{t}^+(M,n)=0$, detailed balance
would be fulfilled for all $m$, and for long times
$P(m,n,t)$ would reach the stationary density:
  \be
P^{\rm st}(m,n)=\mathscr{Z}(m,n)/\mathscr{Z}\label{eq:P_st}
  \ee
where $\mathscr{Z}=\sum_{m,n} \mathscr{Z}(m,n)$. Here, in contrast,
our absorbing condition provides that $P(m,n,t\rightarrow \infty)=0$,
and $P(m,n,t)$ is in fact an improper probability distribution. An
explicit expression for $\mathscr{Z}(m,n)$ and for physically
realistic transfer coefficients, satisfying the detailed balance
conditions above, are given in section \ref{sec:partition_function}.

The general solution to the master equation (\ref{master}) can be decomposed
into the eigenmodes according to
\begin{equation}
\label{eigenex}
P(m,n,t)=P(m,n,t|m',n')=\sum_p c_p(m',n') Q_p(m,n)e^{-\eta_pt},
\end{equation}
where the expansion coefficients $c_p(m',n')$ are determined by the
initial condition.  Inserting above expansion into equation
(\ref{master}) produces the eigenvalue equation
\begin{eqnarray}
\nonumber
& &\mathsf{t}^+(m-1,n)Q_p(m-1,n)+\mathsf{t}^-(m+1,n)Q_p(m+1,n)\\
\nonumber
&&\hspace{0.8cm}-\left[\mathsf{t}^+(m,n)+\mathsf{t}^-(m,n)\right]Q_p(m,n)\\
\nonumber
&&+\mathsf{r}^+(m,n-1)Q_p(m,n-1)+\mathsf{r}^-(m,n+1) Q_p(m,n+1)\\
&&\hspace{0.8cm}-\left[\mathsf{r}^+(m,n)+\mathsf{r}^-(m,n)\right]Q_p(m,n)=-\eta_pQ_p(m,n)
\label{eigen1}
\end{eqnarray}
for the $p$th eigenmode, with eigenvalues $\eta_p$ and eigenfunctions
$Q_p(m,n)$. We label the eigenvalues such that $0<\eta_0<\eta_1<\eta_2
\ \ldots <\eta_M$. We note that all $\eta_p$ are real and positive
(see \ref{proof} and \cite{gardiner,risken}) which guarantees that
$P(m,n,t\rightarrow \infty)=0$, as it should for an ergodic system
with a sink (see equation \eref{absorb}). The eigenvectors $Q_p(m,n)$
satisfy the orthonormality relation given in equation
\eref{eq:ortho_relation}. We prefer using an eigenvalue approach
(spectral representation) to the present problem rather than solving
the master equation in real time, since the eigenvalue approach avoids
time discretization problems.

\section{Mean first passage time of translocation}\label{passage_time}

We here derive an expression for the mean first passage time of translocation,
i.e., the mean time it takes the biopolymer, while being driven by chaperone
(un)binding, to fully cross the membrane pore. We will determine the mean
first passage time as a function of the eigenvalues $\eta_p$ and the
eigenfunctions $Q_p(m,n)$ defined through equation \eref{eigen1}.

As before, denote by $P(m,n,t|m',n')$ the conditional probability to find the
system in state $(m,n)$ at time $t$, given the initial condition $(m',n')$
at time $t=0$. Then the  (survival) probability that the absorbing boundary
at $m=M+1$ has not yet been reached up to time $t$ is
\begin{equation}
\mathscr{S}(m',n',t)=\sum_{m=0}^M\sum_{n=0}^{n^{\mathrm{max}}(m)}
P(m,n,t|m',n').
\end{equation}
The probability that the absorbing boundary is reached within the time
interval $[t,t+dt]$ is
\begin{eqnarray}
\nonumber
f(m',n',t)dt&=&\mathscr{S}(m',n',t)-\mathscr{S}(m',n',t+dt)\\
&=&-\left(\frac{\partial}{\partial t}\mathscr{S}(m',n',t)\right)dt.
\end{eqnarray}
This expression is positive, as $\mathscr{S}$ is decreasing with time.
From the first passage time density $f(m',n',t)$, the mean first passage
time thus follows by integration:
\begin{equation}
\label{mfpt}
\intercal(m',n')=\int_0^{\infty}tf(m',n',t)dt=\int_0^{\infty}\mathscr{S}(m',n',t)dt.
\end{equation}
To express the mean first passage time $\intercal$ in terms of the eigenvalues and
eigenfunctions, we introduce the eigenmode expansion (\ref{eigenex}) into
equation (\ref{mfpt}), yielding
\begin{eqnarray}
\nonumber
\intercal (m',n')&=&\sum_{m,n}\int_0^{\infty}\sum_p c_p(m',n')Q_p(m,n)e^{-\eta_pt}dt\\
&=&\sum_p\eta_p^{-1}\frac{Q_p(m',n')}{P^{st}(m',n')}\sum_{m,n}Q_p(m,n).\label{eq:T}
\end{eqnarray}
where $P^{st}(m,n)$ is given in equation \eref{eq:P_st}.  We have
above made use of the orthonormality relation \eref{eq:ortho_relation}
in order to express $c_p(m',n')$ in terms of the initial probability
density $P(m,n,0|m',n')$, and used the fact that this general
initial condition takes the explicit form
$P(m,n,0|m',n')=\delta_{m,m'}\delta_{n,n'}$. Equation \eref{eq:T} is
the discrete counterpart of the continuous result derived in
\cite{gardiner}, and expresses the mean first passage time (for any
given initial condition, specified by $m'$ and $n'$) in terms of the
eigenvalues and eigenvectors of equation \eref{eigen1}.

Using the general result \eref{eq:T} together with the special initial
condition \eref{initial} considered here,  we see that
the mean translocation time for a polymer starting with its head in
the pore becomes:
  \begin{equation}
\intercal =\intercal (0,0)=\sum_p \intercal_p \eta_p^{-1}, \label{eq:T2}
  \end{equation}
with
\begin{equation}
\intercal_p=\frac{Q_p(0,0)}{P^{\mathrm{st}}(0,0)}\sum_{m,n}Q_p(m,n).\label{eq:T_p}
\end{equation}
The mean translocation time is thus obtained by summing up the
relaxation times $\eta_p^{-1}$ with weight-factors $\intercal_p$, 
in turn being determined by the eigenfunctions $Q_p(m,n)$.

\section{Partition function and transfer rates}\label{sec:partition_function}

The partition coefficient $\mathscr{Z}(m,n)$ appearing in the detailed
balance conditions \eref{dbt} and \eref{dbr} constrain, but do not
fully specify, the transfer coefficients $\mathsf{t}^\pm(m,n)$ and
$\mathsf{r}^\pm(m,n)$. In this section we give explicit expressions for
both the partition coefficient as well for the transfer coefficients.

For a given value of translocation coordinate $m$ and a given number of bound chaperones $n$ the partition coefficient is:
\begin{equation}
\mathscr{Z}(m,n)=\mathscr{Z}^{\mathrm{poly}}(m)\mathscr{Z}^{\mathrm{chap}}(m,n),
\end{equation}
corresponding to the product of the partition coefficients
$\mathscr{Z}^{ \mathrm{poly}}(m)$ of the undressed biopolymer and
$\mathscr{Z}^{\mathrm{chap}} (m,n)$ of the chaperone (un)binding,
which in turn depends on the translocation coordinate $m$ defining the
number of accessible binding sites, see figure \ref{fig:scheme}. The
polymeric factor $\mathscr{Z}^{\mathrm{poly}}(m)$ measures the
accessible degrees of freedom (or better, their confinement) of the
biopolymer chain threading through the pore. For a flexible chain,
  this would invoke the critical exponent for a polymer grafted to a
  surface \cite{muthu}. Single-stranded DNA or RNA would qualify for
  such a description. Double-stranded DNA and proteins, are more of a
  semiflexible nature, and corrections to the scaling behaviour would
  ensue.  Moreover, the interactions between chain and pore are not
fully known, compare the discussions in references
\cite{bates,flomenbom1}. For simplicity, and to solely concentrate on
the chaperone effects, we assume our chain to be completely stiff, so
that
  \be
\mathscr{Z}^{\mathrm{poly}}(m)=1.
  \ee
Note, however, that any expression for
$\mathscr{Z}^{\mathrm{poly}}(m)$ can easily be implemented through the
numerical formalism discussed below, provided that the chaperone
binding does not depend on the curvature of the polymer.

The second factor in the partition function $\mathscr{Z}(m,n)$, the
contribution $\mathscr{Z}^{\mathrm{chap}}(m,n)$ from chaperone (un)binding,
\begin{equation}
\mathscr{Z}^{\mathrm{chap}}(m,n)=\kappa^n\Omega^{\mathrm{chap}}(m,n)
\end{equation}
combines the $n$-fold gain of the binding energy, entering through the factor
$\kappa^n$, with the number of possible ways of arranging $n$ proteins along
$m$ biopolymer monomers,
\begin{equation}
\Omega^{\mathrm{chap}}(m,n)={m-(\lambda-1)n \choose n}=\frac{(m-[\lambda
-1]n)!}{n!(m-\lambda n)!},
\end{equation}
compare references \cite{McQuistan,Epstein,ssb1,ambme}.
Here, we also introduced the binding strength 
  \be
\kappa=c_0 K^{\mathrm{eq}},\label{eq:kappa}
  \ee
that, in turn, depends on the chaperone concentration $c_0$ in the
surrounding solution and the equilibrium binding constant
$K^{\mathrm{eq}}=v_0\exp \left(|E_{\mathrm{bind}}|/k_BT\right)$ with
the typical volume $v_0$ of the chaperones; $E_{\mathrm{bind}}$ is the
chaperone binding energy, $\beta=1/k_B T$ with $k_B$ being the
Boltzmann constant and $T$ the temperature.

The transfer rates $\mathsf{t}^{\pm}(m,n)$ and $\mathsf{r}^{\pm}(m,n)$
governing the transitions $m\to m\pm 1$ and $n\to n\pm 1$ in the
master equation (\ref{master}) are subject to the detailed balance
conditions (\ref{dbt}) and (\ref{dbr}). Here, we seek concrete
expressions for the rates. For the forward translocation rate
$\mathsf{t}^{\pm}$, we solely observe the influence of the polymeric
degrees of freedom. We assume that the rate limiting step for the
polymer motion originates from interactions in the pore
\cite{lubensky} and choose the form
\begin{equation}
\mathsf{t}^+(m,n)=k\frac{\mathscr{Z}^{\mathrm{poly}}(m+1)}{
\mathscr{Z}^{\mathrm{poly}}(m)}=k=\mathsf{t}^+(m).\label{eq:t_plus}
\end{equation}
 It simplifies to the constant rate $k$ due to the assumption of a
stiff chain, $\mathscr{Z}^{\mathrm{poly}}(m)=1$, where $k$ is the rate
for making a diffusive jump of length $\sigma$, see introduction and
figure \ref{fig:scheme}.  Conversely, for the backward rate, we
specify
\begin{equation}
\mathsf{t}^-(m,n)=k\frac{\mathscr{Z}(m-1,n)}{\mathscr{Z}(m,n)}=k\frac{\mathscr{Z}^{\mathrm{chap}}(m-1,n)}{\mathscr{Z}^
{\mathrm{chap}}(m,n)}=k\frac{\Omega^{\mathrm{chap}}(m-1,n)}{\Omega^{
\mathrm{chap}}(m,n)}.\label{eq:t_minus}
\end{equation}
In the case of the $\mathsf{t}^{\pm}$, we thus choose that while the
forward rate is independent of the number of bound chaperones and
their configurations, the sliding of the chain toward the cis side A of the
membrane will, in general, be opposed by bound proteins,
$\mathsf{t}^-(m,n)\le k$. We have chosen $\mathsf{t}^-$ such that it
is proportional to the probability
$\Omega^{\mathrm{chap}}(m-1,n)/\Omega^{ \mathrm{chap}}(m,n)$ that the
binding site closest to the pore on side B is vacant. In the limit
$n=0$, both forward and backward rates are identical:
$\mathsf{t}^+(m,n=0)=\mathsf{t}^- (m,n=0)=k$, and we have pure,
unbiased diffusion with a bare diffusion constant $D_0=\sigma^2 k$,
\footnote{The bare diffusion constant $D_0$ should not be confused
  with the effective diffusion constant relevant for general driven
  motion over distances much larger than $\sigma$, see reference
  \cite{lubensky}.}  as it should be. We point out that whenever the
polymer is not fully occupied, the binding proteins may, in general,
slide along the DNA \cite{Harada}, thereby possibly exploring the
different bound chaperone configurations on a faster timescale. Such
sliding motion was present in the Brownian simulations in reference
\cite{zandi}.

For the transfer coefficients associated with chaperone (un)binding we
specify:
\begin{eqnarray}
\nonumber
\mathsf{r}^+(m,n)&=&(n+1)q\frac{\mathscr{Z}^{\mathrm{chap}}(m,n+1)}{\mathscr{Z}
^{\mathrm{chap}}(m,n)}\label{eq:r_plus}\\
&=&k \gamma \kappa\frac{(n+1) \Omega^{\mathrm{chap}}(m,n+1)}{\Omega^{\mathrm{chap}}
(m,n)};\\
\mathsf{r}^-(m,n)&=&nq=k \gamma n.\label{eq:r_minus}
\end{eqnarray}
To make comparison with the chaperone unbinding rate $q$ easier, we
introduced here the dimensionless ratio 
\begin{equation}
\gamma=\frac{q}{k}\label{eq:gamma}
\end{equation} 
between the the unbinding rate $q$ of a single chaperone (for $n=1$ we
have $\mathsf{r}^-=q$) and the diffusive rate $k$. We have chosen
$\mathsf{r}^+$ to be proportional to the number of ways
$\mathscr{N}=(n+1)
\Omega^{\mathrm{chap}}(m,n+1)/\Omega^{\mathrm{chap}}(m,n)$ of adding
one additional chaperone if there are already $n$ bound. In the
general case $\mathscr{N}$ includes a ``car parking effect''
\cite{McQuistan,ambme}, i.e., the fact that a chaperone can only bind
in between two chaperones which are separated by a distance equal to
or larger than the chaperone size.  In the limit $\lambda=1$ we have
independent binding and the resulting expression becomes
$\mathscr{N}=m-n$ as it should, and $\mathsf{r}^+$ is thus
proportional to the number of free binding sites, compare the
discussion in \cite{ssb1,vankampen}. We note that $\mathsf{r}^+$
depends linearly on the volume concentration $c_0$ of chaperones
(compare reference \cite{zandi}) through the parameter $\kappa=c_0
K^{\rm eq}$, where $K^{\rm eq}$ is the binding constant as
before. Conversely, the unbinding rate is simply proportional to the
number of chaperones bound to the biopolymer (as it should
\cite{vankampen}) and independent of the concentration of chaperones. A
word on the choice of this asymmetric form is in order. To put the
full effect of chaperone concentration and binding energy contained in
$\mathscr{Z}^{\mathrm{chap}}$ into the binding rate $\mathsf{r}^+$,
and therefore leave the unbinding rate $\mathsf{r}^-$ unaffected
corresponds to the physical picture that the unbinding process should
solely depend on the dissociation of local bonds. It should also be
noted that cooperative binding of chaperones can in principle be
included in our formalism \cite{ambme}.

The explicit expressions \eref{eq:t_plus}, \eref{eq:t_minus},
\eref{eq:r_plus} and \eref{eq:r_minus} for the transfer coefficients,
together with the boundary conditions in equations \eref{eq:bc1},
\eref{eq:bc2}, \eref{eq:bc3} and \eref{eq:bc4} completely determine
the translocation/chaperone (un)binding dynamics through the master
equation \eref{master} or the associated eigenvalue equation
\eref{eigen1}. In the next section we solve these equations in order
to obtain the mean translocation time, using the expression
\eref{eq:T2} derived in the previous section.

\section{General and limiting results}\label{sec:results}

\begin{figure}
\begin{center}
\scalebox{0.5}{\epsfbox{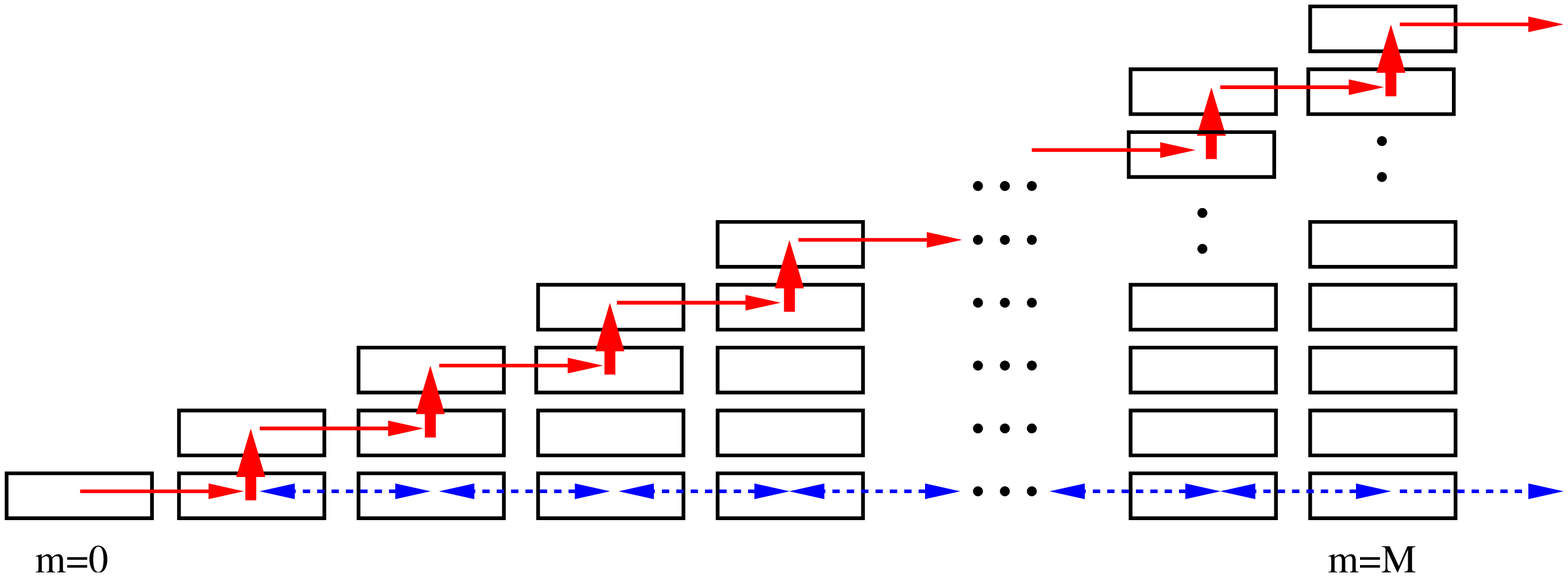}}
\end{center}
\vspace{0.5cm}
\begin{center}
\scalebox{0.8}{\epsfbox{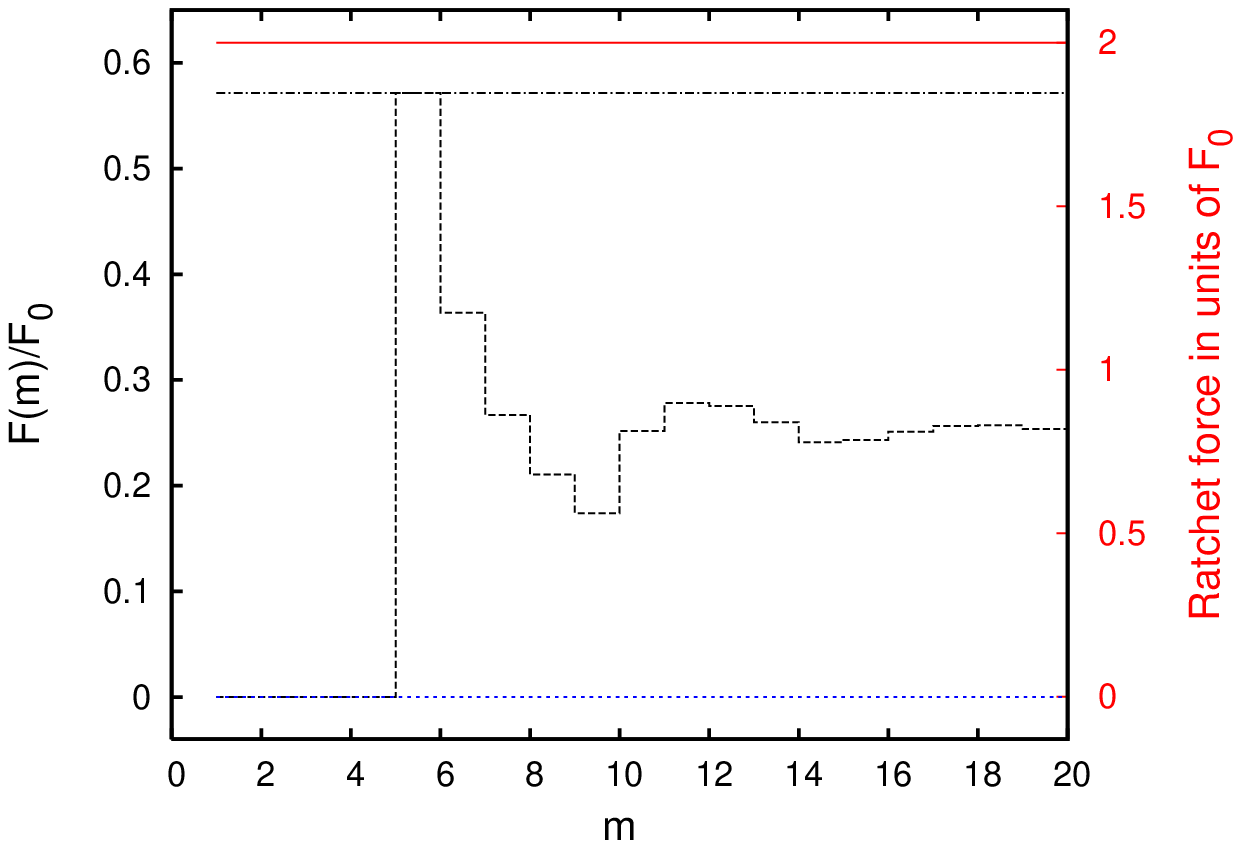}}
\end{center}
\caption{Top: Schematic of the analytically tractable limiting cases
  of the coupled translocation-chaperone (un)binding dynamics. (i) The
  dashed blue arrows indicate the purely, unbiased diffusive motion
  for slow binding. (ii) The solid red arrows indicate the
  translocation dynamics for ratchet motion in the case of fast
  chaperone binding. Notice that for these two cases only a small part
  of the phase space is explored before the polymer is absorbed at
  $m=M+1$. In contrast, for (iii) the fast (un)binding case, the
  entire phase space is explored before absorption. Bottom: Effective
  force $F(m)$ for the three limiting regimes above. For (i) diffusive
  motion we have $F(m) \equiv 0$ (blue dashed line); for (ii) ratchet
  motion we effectively have $F(m)=2F_0$ (red solid line referring to
  the right ordinate), with a
  characteristic force $F_0=k_B T/\sigma$; for (iii) fast (un)binding
  we have $F(m)=2F_0 (\tilde{\mathsf{t}}^+ (m)-\tilde{\mathsf{t}}^-
  (m))/(\tilde{\mathsf{t}}^+ (m)+\tilde{\mathsf{t}}^- (m))$ where
  $\tilde{\mathsf{t}}^\pm (m)$ are defined in subsection
  \ref{fast_bind}. We here use a binding strength $\kappa=0.8$, and
  two different chaperone sizes $\lambda=1$ (the upper dash-dotted line) and
  $\lambda=5$ (the lower dashed curve). Note that the onset of a non-zero
  force is at $m=\lambda$, i.e. the force is zero unless there are
  sufficiently many monomers to accommodate at least one
  chaperone. Also notice the oscillations in the force for the case
  $\lambda=5$. We point out that the forces above are not explicitly
  used in the dynamical scheme, but are convenient for illustrative
  purposes.}
\label{lattice_jumps}
\end{figure}

In this section we obtain results for the mean translocation time for
different physical parameters. In particular three cases can be
distinguished: (i) slow chaperone binding leading to unbiased pure
diffusion for a finite translocating chain; (ii) fast chaperone
binding but slow unbinding leading to a pure chaperone-mediated
ratchet motion; and (iii) fast chaperone (un)binding dynamics
tractable via adiabatic elimination, leading to a one variable
effective force, compare figure \ref{lattice_jumps}. This analytical
treatment is complemented with (iv) numerical analysis. We stop to
note that the emergence of the three limiting cases from a general
scheme is the major advantage of the approach pursued herein in
comparison to the one-dimensional treatment in reference \cite{ambme},
which explicitly accounts for case (iii) above only.

\subsection{Slow binding dynamics, $\gamma \kappa \rightarrow 0$}\label{slow_bind}

Slow chaperone binding is prevalent when the dimensionless ratio
$\gamma\kappa \rightarrow 0$, as in this limit we have that
$\mathsf{r}^+(m,n)\to 0$ (see equation
\eref{eq:r_plus}). Experimentally one can reach small values of
$\kappa$ and hence the slow binding regime, by sufficiently low
chaperone concentration $c_0$ or for small binding energy
$E_{\mathrm{bind}}$, see equation \eref{eq:kappa}. As initially $n=0$
(see equation \eref{initial}), the dynamics is fully described by the
transfer coefficients
\begin{equation}
\mathsf{t}^+(m,0)=\mathsf{t}^-(m,0)=k, \label{eq:tm0}
\end{equation}
together with the reflecting boundary (compare equation (\ref{reflect}))
\begin{equation}
\mathsf{t}^-(0,0)=0
\end{equation}
and the absorbing boundary condition (see equation (\ref{absorb}))
\begin{equation}
\mathsf{t}^-(M+1,0)=0.
\end{equation}
We see from equation \eref{eq:tm0} that in the slow binding limit the
rate for a forward step equals the backward rate and therefore the
motion is purely diffusive for a finite chain. Making use of the
general expression for the mean first passage time for one-dimensional
motion contained in equation \eref{eq:T_1d} we arrive at
\begin{equation}
\intercal =\intercal_{\rm diff}=\sum_{m=0}^M\sum_{m'=0}^m\ k^{-1}=k^{-1}\left[(M+1)
\left(1+\frac{M}{2}\right)\right].\label{eq:T_slow}
\end{equation}
For large $M$, the mean first passage time thus scales like
$\intercal\simeq M^2/2k$, as it should for a purely diffusive first
passage process. In terms of the bare diffusion constant
$D_0=\sigma^2k$ and the chain length $L=M\sigma$ we have
$\intercal\simeq L^2/2D_0$ in agreement with
\cite{simon,peskin,zandi}. The diffusive motion on the general
configuration lattice is schematically illustrated by the blue dashed
arrows in figure \ref{lattice_jumps} (top). Notice that due to the
slow binding only a small part of the phase space is explored before
the polymer is absorbed. It is sometimes useful to picture the
dynamics as jump motion in a force field; for purely diffusive motion
we have a force $F(m) \equiv 0$, i.e. the force is zero, as
schematically illustrated in figure \ref{lattice_jumps} (bottom). In
figure \ref{fig:tau_M} the mean translocation time \eref{eq:T_slow} is
shown and compared to the different cases discussed in the remaining
part of this section.

\subsection{Fast binding but slow unbinding, $\gamma\to 0$, $\gamma\kappa\to\infty$ and $\lambda=1$: ratchet motion}\label{ratchet}

As the effective rate $\gamma$ and the binding strength can be chosen
independently, we can specify them such that we encounter slow
unbinding designated by $\gamma\to 0$ but fast binding,
$\gamma\kappa\to\infty$.  For small chaperones, $\lambda=1$ (univalent
binding), this corresponds to the stepwise transitions marked by the
red arrows in figure \ref{lattice_jumps}. Each time a monomer of the
biopolymer exits from the membrane pore towards the trans side, it is
immediately occupied by a chaperone. The irreversibly bound chaperones
do not unbind and prevent backwards motion (see equation
\eref{eq:bc2}) so that the motion becomes unidirectional with
effective rates $\overline{\mathsf{t}}^+(m)=k$ and
$\overline{\mathsf{t}}^-(m)=0$. The general master equation
\eref{master} then reduces to the effective (1+1)-variable equation
\begin{equation}
\frac{\partial\overline{P}(m)}{\partial t}=k\left[\overline{P}(m-1,t)-
\overline{P}(m,t)\right],\label{eq:shot_noise}
\end{equation}
where we introduced the short-hand notation
$\overline{P}(m,t)=P(m,n^{\rm max}(m),t)$.  The process is that of
taking $M+1$ forward steps with a rate per step $k$, and the mean
first passage time therefore becomes
\begin{equation}
\intercal =\intercal_{\rm ratchet}= k^{-1}(M+1).\label{eq:T_ratchet}
\end{equation}
Thus, $\intercal$ scales linearly with $M$ for large $M$ in this
ratchet limit. The mean translocation time \eref{eq:T_ratchet} is
shown in figure \ref{fig:tau_M}. Equations of the same form as
\eref{eq:shot_noise} also occur in the theoretical description of shot
noise, and correspond to the forward mode of the wave equation. For
ratchet motion we define an effective force $F(m)=2F_0
(\overline{\mathsf{t}}^+(m)-\overline{\mathsf{t}}^-(m))/(\overline{\mathsf{t}}^+(m)+\overline{\mathsf{t}}^-(m))=2F_0$,
where $F_0=k_BT/\sigma$; the effective force associated with ratchet
motion is thus constant, as schematically shown in figure
\ref{lattice_jumps} (bottom).

Comparing equations \eref{eq:T_slow} and \eref{eq:T_ratchet} we see
that $\intercal_{\rm ratchet}/\intercal_{\rm diff}=2/M $ for large
$M$. This results contrasts with the results of a Brownian ratchet
\cite{simon,peskin} for which the same quantity equals $1/M $. This
difference originates from the fact that here the size of a chaperone
equals the diffusive step length $\sigma$, whereas in the Brownian
ratchet $\sigma\rightarrow 0$ (continuous diffusion versus our
stepwise diffusion), but with a finite chaperone size.
\footnote{The results from \cite{simon,peskin} can be recovered if we imagine 
the chaperones in the ratchet case to be fast binding but immobile after 
binding. We would then expect the mean first passage time for $M\gg\lambda\gg 
1$ to be the sum of $M/\lambda$ diffusive mean first passage times   
$k^{-1}\lambda^2/2$. Thus the ratio with the purely diffusive mean first 
passage time in equation \eref{eq:T_slow} would become $\intercal_{\rm 
ratchet}/\intercal_{\rm diff}=\lambda/M $ in agreement with 
\cite{simon,peskin}.}

\subsection{Fast (un)binding dynamics, $\gamma\rightarrow \infty$ and $\gamma\kappa \rightarrow \infty$}\label{fast_bind_and_unbind}\label{fast_bind}

Finally, we address the case of fast binding, $\gamma\kappa\to\infty$, and
fast unbinding, $\gamma\to\infty$. Under these conditions, it is possible
to adiabatically eliminate the fast variable $n$ \cite{ssb1,risken}, resulting
in the effective forward and backward rates
\begin{equation}
\overline{\mathsf{t}}^{\pm}(m)=\sum_{n=0}^{n^{\mathrm{max}}(m)}\mathsf{t}^{\pm}
(m,n)\frac{\mathscr{Z}(m,n)}{\mathscr{Z}(m)}.
\end{equation}
For a stiff polymer, implying $\mathsf{t}^+(m,n)=k$ (see equation
\eref{eq:t_plus}), we obtain the following simple expression for the
forward rate: $\overline{\mathsf{t}}^+(m)=k$. When calculating the
backward rate $\overline{\mathsf{t}}^-(m)$ one should remember to
incorporate the boundary rates as given in equation
\eref{eq:bc2}. Here, in general,
$\mathscr{Z}(m)=\mathscr{Z}^{\mathrm{poly}}(m)\mathscr{Z}^{\mathrm{chap}}(m)$,
and
\begin{equation}
\mathscr{Z}^{\mathrm{chap}}(m)=\sum_{n=0}^{n^{\mathrm{max}}(m)}
\mathscr{Z}^{\mathrm{chap}}(m,n).
\end{equation}
Making use of the fact that the transfer coefficients satisfy the
detailed balance condition:
$\overline{\mathsf{t}}^+(m-1)\mathscr{Z}(m-1)=\overline{\mathsf{t}}^-(m)\mathscr{Z}(m)$
(see reference \cite{ssb1}) so that we can use equation
\eref{eq:T_1d_alt}, we obtain the mean first passage time
  \be
\intercal =\sum_{m=0}^M \frac{1}{\overline{\mathsf{t}}^+(m)\mathscr{Z}(m)} \sum_{m'=0}^m \mathscr{Z}(m')=k^{-1} \sum_{m=0}^M \frac{1}{\mathscr{Z}(m)} \sum_{m'=0}^m \mathscr{Z}(m'),\label{eq:T_fast}
  \ee
where we have used that here $\overline{\mathsf{t}}^+(m)=k$. The sum
in \eref{eq:T_fast} can straightforwardly be numerically evaluated
and the resulting mean translocation time is shown in figure
\ref{fig:tau_M}.

In order to simplify the mean translocation time
\eref{eq:T_fast} further for large chain lengths $M$, a knowledge of
the partition function $\mathscr{Z}(m)$ for large $m$ suffices. Instead of the combinatorial approach given in the
previous subsections, the partition function for large $m$ is
obtained using the approach in reference \cite{DiCera_Kong} (see
also \cite{Kong}), giving 
  \be
\mathscr{Z}(m)\simeq C \Lambda_{\rm max}^m\label{eq:Z_DiCera},
  \ee
where $C$ is a constant and (assuming no cooperativity effects
\cite{DiCera_Kong,ambme}) $\Lambda_{\rm max}$ is the largest root
$\Lambda$ to the algebraic equation
  \be
\Lambda^{\lambda}-\Lambda^{\lambda-1}-\kappa =0.\label{eq:algebraic_eq}
  \ee
The equation above is of order $\lambda$, i.e., the order is determined
by the size of the chaperones. Introducing equation \eref{eq:Z_DiCera} in
equation \eref{eq:T_fast} we find that the mean first passage time for
very large $M$ approaches
  \be
\intercal |_{M\gg 1}\rightarrow \intercal_{\rm ad}= k^{-1} \frac{\Lambda_{\rm max}}{\Lambda_{\rm max}-1} (M+1),\label{eq:tau_M_large}
  \ee
which proves that for large chain lengths the adiabatic result for
$\intercal$ scales as $M$, just as for ratchet dynamics, but in
general with a smaller prefactor than for ratchet motion.  For the
case of weak binding, $\kappa$ small, so that $\ln \Lambda_{\rm
max}\ll 1$ we can write equation \eref{eq:tau_M_large} according to
$\intercal \approx k^{-1} \ln \Lambda_{\rm max}^{-1}(M+1)$, so that
$\intercal\propto \ln \Lambda_{\rm max}^{-1}$ in agreement with the
results in reference \cite{ambme}.

For univalent binding ($\lambda=1$) we solve equation
\eref{eq:algebraic_eq} and find the explicit result $\Lambda_{\rm
max}=(1+\kappa)$. Interestingly, the mean translocation time is then
inversely proportional, $\intercal \propto 1/f$, to the filling fraction,
$f=\kappa/(1+\kappa)$, of the chain, see appendix C in reference
\cite{ambme}; since $0\le f\le 1$ the translocation process is slower
than for ratchet motion as it should be, compare equation
\eref{eq:T_ratchet}. We note that $\intercal_{\rm ad}/\intercal_{\rm
diff}=2(1+\kappa^{-1})/M$, see equations \eref{eq:T_slow} and
\eref{eq:tau_M_large}.
\footnote{The corresponding result for a Brownian ratchet is
 $\intercal_{\rm ad}/\intercal_{\rm diff}=(1+2\kappa^{-1})/M$. The
 difference has the same origin as discussed in subsection
 \ref{ratchet}.}  For divalent binding ($\lambda =2$) equation
 \eref{eq:algebraic_eq} becomes a second order algebraic equation,
 which can straightforwardly be solved, yielding $\Lambda_{\rm
 max}=[1+ (1+4\kappa)^{1/2}]/2$.  In figure \ref{fig:tau_M} 
 the result as contained in \eref{eq:tau_M_large} is shown for the
 cases $\lambda=1$ and $\lambda=5$; for the case $\lambda=5$ the
 algebraic equation \eref{eq:algebraic_eq} is solved numerically.

We point out that, in contrast to the cases of slow binding and
ratchet motion discussed in the previous subsections, for the fast
binding and unbinding case (reversible binding \cite{ambme})
considered here typically the larger part of the phase space is
explored, see figure \ref{lattice_jumps} (top), before absorption. For
fast (un)binding we define the effective force $F(m)=2F_0
(\tilde{\mathsf{t}}^+ (m)-\tilde{\mathsf{t}}^-
(m))/(\tilde{\mathsf{t}}^+ (m)+\tilde{\mathsf{t}}^- (m))$, with
$F_0=k_BT/\sigma$ as before. The force is illustrated in figure
\ref{lattice_jumps} (bottom), showing a constant value for large $m$;
this asymptotic value is smaller than that for ratchet motion. Also,
notice that for $\lambda=5$ the force oscillates with a period
$\approx \lambda$ for smaller $m$ in agreement with reference \cite{ambme}.

\subsection{General case, numerical analysis}

\begin{figure}
\begin{center}
\scalebox{0.55}{\epsfbox{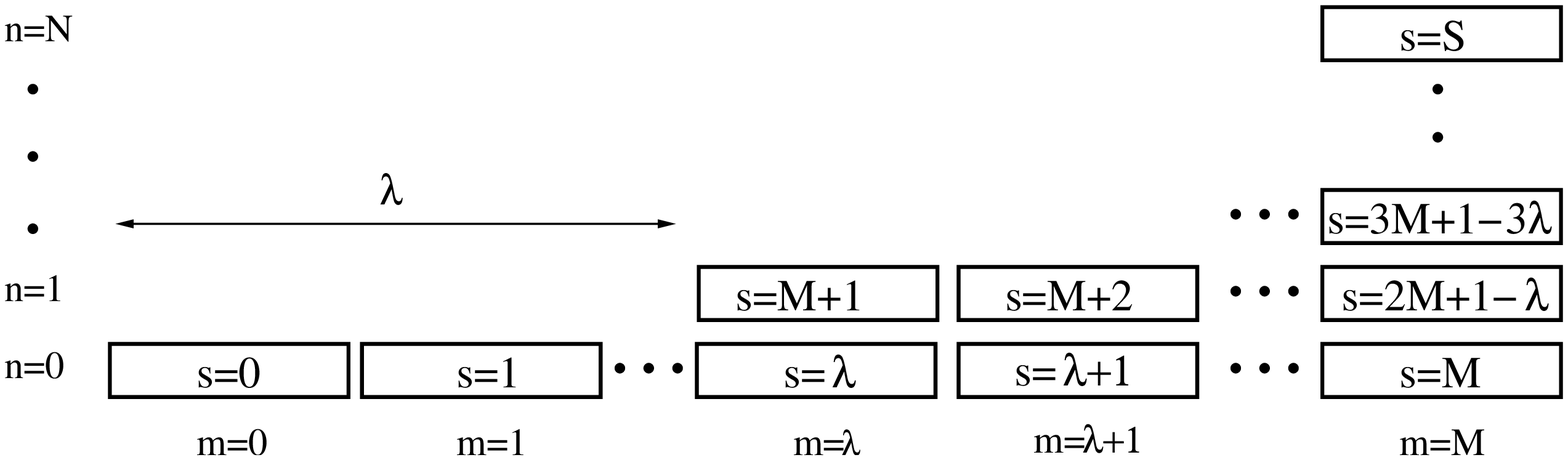}}
\end{center}
\caption{Enumeration scheme for the numerical analysis: The two-dimensional
grid points $(m,n)$ are replaced by a one-dimensional running variable $s$.
See text for details.}
\label{fig:s_space}
\end{figure}

 To solve the eigenvalue equation (\ref{eigen1}) by a numerical
scheme, it is convenient to replace the two-dimensional grid points $(m,n)$
by a one-dimensional coordinate $s$ counting all lattice points, compare
\cite{ssb1} and \cite{recipe}. We choose the enumeration illustrated in
figure \ref{fig:s_space}. 
From this figure we notice that $n\in [0,N]$ (where
$N=[M/\lambda]$) and that $m\in [n\lambda,M]$. An arbitrary
$s$-point can be obtained from a specific $(m,n)$ according to:
  \be
s=n(M+1)-\sum_{n'=1}^n n'\lambda +m = n(M+1)-\frac{n(n+1)}{2}\lambda+m.\label{eq:s}
  \ee
From this relation we notice that the maximum $s$ value is
  \be 
S={\rm max}\{ s\}=(N+1)\left(M+1-\frac{N\lambda}{2}\right)-1.
  \ee
Expression \eref{eq:s} allows us to change the transfer coefficients
to the $s$-variable, $\mathsf{t}^\pm (m,n)\rightarrow \mathsf{t}^\pm
(s)$ and $\mathsf{r}^\pm (m,n)\rightarrow \mathsf{r}^\pm (s)$, using
the explicit expressions \eref{eq:t_plus}, \eref{eq:t_minus},
\eref{eq:r_plus} \eref{eq:r_minus} for the transfer coefficients,
together with the boundary conditions in equations \eref{eq:bc1},
\eref{eq:bc2}, \eref{eq:bc3} and \eref{eq:bc4}. From equation
\eref{eq:s} and figure \ref{fig:s_space} we notice that a local jump
in the $m$-direction corresponds to a local jump also in $s$-space,
i.e., that
  \bea
m\rightarrow m-1 &\Longleftrightarrow & s\rightarrow s-1 |_{\rm 1 \le m \le M}\ ,\nonumber\\
m\rightarrow m+1 &\Longleftrightarrow & s\rightarrow s+1 |_{\rm 0\le m \le M-1}\ .
  \eea
However a jump in the $n$-direction is equal to a non-local jump in $s$-space:
  \bea
n\rightarrow n-1 &\Longleftrightarrow & s\rightarrow s-\triangle s^- |_{\rm 1 \le n\le n^{\rm max}(m)}\ ,\nonumber\\
n\rightarrow n+1 &\Longleftrightarrow & s\rightarrow s+\triangle s^+|_{\rm 0 \le n\le n^{\rm max}(m)-1}\ ,
  \eea
with $\triangle s^-=M+1-n\lambda$ and $\triangle
s^+=M+1-(n+1)\lambda$. Using the results above, we find that the eigenvalue
problem \eref{eigen1} can be written in matrix form as
  \be
\sum_{s'}W(s,s')Q_p(s')=-\eta_p Q_p(s)\label{eigen2}
  \ee
where explicitly the matrix-elements are
  \bea
W(s,s-1)&=&\mathsf{t}^+(s-1),\ {\rm for} \ s  \pitchfork 1\le m\le M, \nonumber\\
&=&0, \ \ {\rm for} \ s  \pitchfork  m=n\lambda\ \wedge \ n=n^{\rm max}(m)  \nonumber\\
W(s,s+1)&=&\mathsf{t}^-(s+1),\ \  {\rm for} \ s \pitchfork  0\le m\le M-1 \nonumber\\
W(s,s-\triangle s^-)&=&\mathsf{r}^+(s-\triangle s^-), \ \  {\rm for} \ s \pitchfork  1\le n\le n^{\rm max}(m), \nonumber\\
W(s,s+\triangle s^+)&=&\mathsf{r}^+(s+\triangle s^+), \ \  {\rm for} \ s \pitchfork 0\le n\le n^{\rm max}(m)-1, \nonumber\\
W(s,s)&=&-(\mathsf{t}^+(s)+\mathsf{t}^-(s)+\mathsf{r}^+(s)+\mathsf{r}^-(s)),\label{eq:W}
  \eea
and the remaining matrix elements are equal to zero. We have
introduced the notation $s\pitchfork $ with the meaning ``$s$ is to be
taken for''. Notice that (with the present enumeration scheme) the
tridiagonal part of the $W$-matrix is determined by jumps in the
$m$-direction, i.e., by $\mathsf{t}^+$ and $\mathsf{t}^-$, whereas
elements outside of the tridiagonal part are determined by
$\mathsf{r}^+$ and $\mathsf{r}^-$. The problem at hand is that of
determining the eigenvalues and eigenvectors of the
$(S+1)\times(S+1)$-matrix $W$ above. Convenient checks of the
numerical results include: (i) The eigenvalues should be real and
negative (so that $\eta_p>0$); (ii) The eigenvectors should satisfy
the orthonormality relation, equation \eref{ortho}.

\begin{figure}
\begin{center}
\scalebox{0.8}{\epsfbox{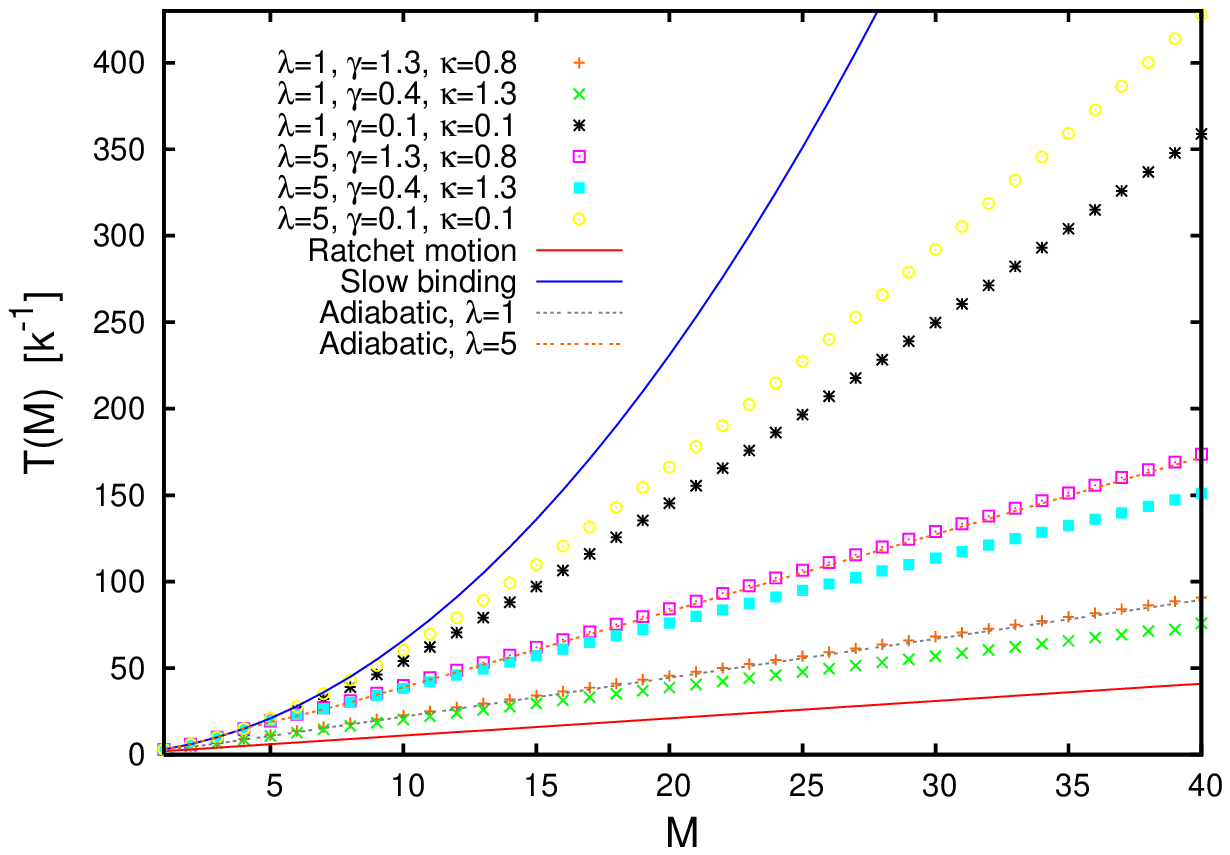}}
\end{center}
\begin{center}
\scalebox{0.8}{\epsfbox{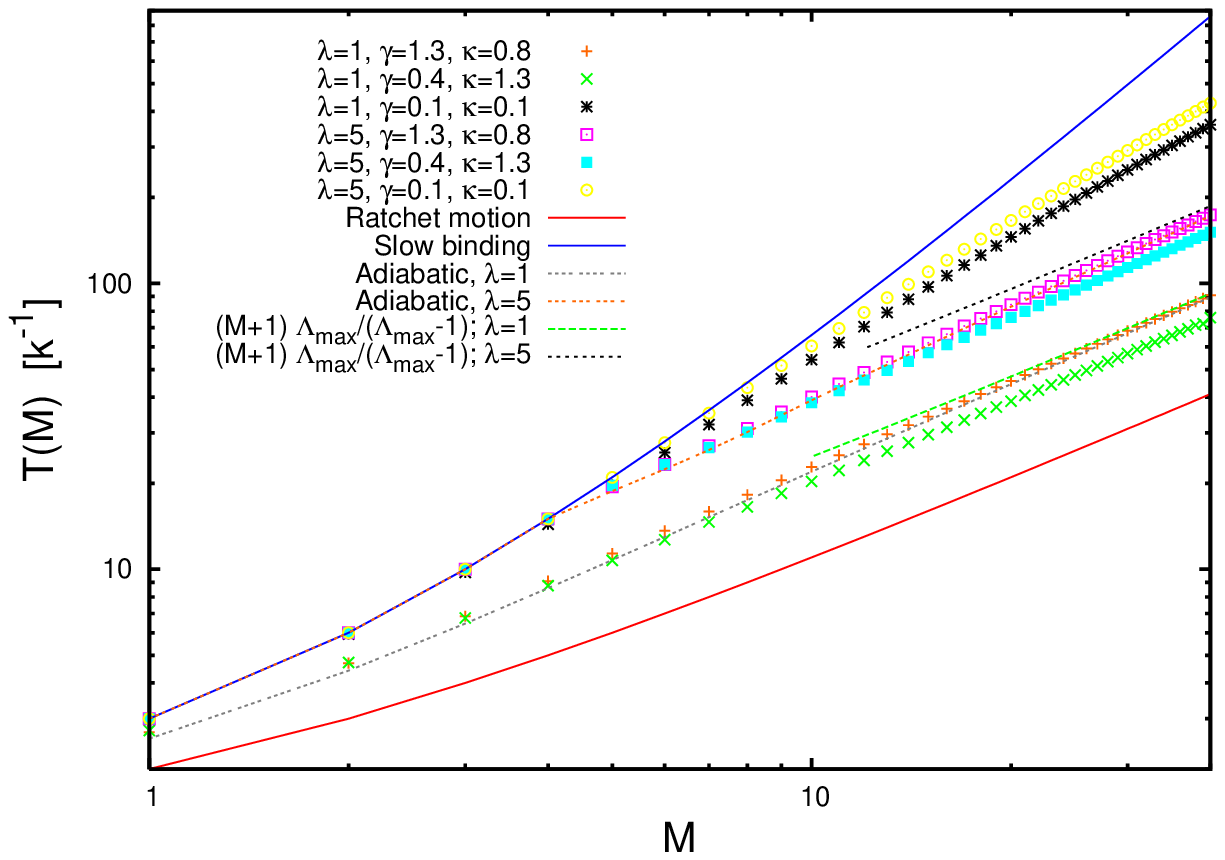}}
\end{center}
\caption{Mean translocation time $\intercal$ as a function of $M$,
using (top) linear axes; (bottom) logarithmic
axes. Results are shown for: (i) slow binding, equation
\eref{eq:T_slow}; (ii) the ratchet result, equation
\eref{eq:T_ratchet}; (iii) fast (un)binding, equation \eref{eq:T_fast}
for different sizes of the chaperones (here $\lambda=1$ and
$\lambda=5$ respectively) with binding strength $\kappa=0.8$. Also, results are shown for (iv) the
general dynamics, as obtained by numerically inverting the transfer
matrix \eref{eq:W} and using the mean first passage time expression
\eref{eq:T_p}. In addition, in the bottom graph the large $M$ result
 \eref{eq:tau_M_large} is shown.}\label{fig:tau_M}
\end{figure}

In figure \ref{fig:tau_M} we show results for the mean translocation
time. From the results of the previous section we see that for ratchet
motion (subsection \ref{ratchet}) as well as for fast binding and
unbinding dynamics (subsection \ref{fast_bind_and_unbind}) the mean
translocation time scales as $\simeq M$. For slow binding as discussed
in subsection \ref{slow_bind} we have that $\intercal $ scales as
$\simeq M^2$ for large $M$. The general results shown in figure
\ref{fig:tau_M} are obtained by numerically solving the eigenvalue
equation \eref{eigen2}, and using the mean first passage time
expression \eref{eq:T_p}. All results are shown for for the case of
small chaperones $\lambda=1$, and for larger chaperones, $\lambda=5$.
We notice the slow binding result \eref{eq:T_slow} and the ratchet
results \eref{eq:T_ratchet} constitute upper and lower limits to the
general mean translocation time $\intercal$. Also, the approach to the
very large $M$ result \eref{eq:tau_M_large} is rather slow, due to
finite size effects.
 
We point out that even though our main focus here has been to calculate the
mean first passage time $\intercal$, any quantity of interest (such
as, for instance, the mean first passage time density $f(t)$, see
section \ref{passage_time}) can be obtained from the solution of the
eigenvalue problem presented in this section.

\section{Conclusions and discussion}

In this study the translocation of a stiff polymer consisting of $M$
monomers through a nanopore in a membrane was investigated in the
presence of binding particles (chaperones) that bind onto the
polymer. Our (2+1)-variable master equation governing the coupling
of the random biopolymer passage and the random chaperone (un)binding
is an extension of a previous continuum Fokker-Planck approach \cite{zandi}.
From our approach three
limiting regimes emerge: (i) For the case of slow binding the motion
is purely diffusive, and the mean translocation time $\intercal$
scales as $\intercal\simeq M^2$ for large $M$; (ii) for fast binding
but slow unbinding, the motion is ratchet-like, for small chaperones
$\lambda=1$, and $\intercal\simeq M$; (iii) for the case of fast
binding and unbinding dynamics, we performed the adiabatic elimination
of the fast variable $n$, and find that for a very long polymer
$\intercal\simeq M$, but with a smaller prefactor than for
ratchet-like dynamics. We point out that our approach is the first, to
our knowledge, that is able to account for (the approach towards)
these three limiting regimes within a single framework. For the
intermediate cases our scheme is solved numerically, which requires
numerically inverting the transfer-matrix $W$ (of approximate size
$M^2\times M^2$), for which an explicit expression is given.

Let us summarize the essential differences between our approach
compared to the original Brownian ratchet ideas in
reference \cite{simon,peskin}. (1) In this reference the polymer
motion through the pore was characterized by continuous diffusion. In
contrast, we assume that the motion proceeds through random jumps of
finite step length, as appropriate, for instance, for a polymer moving
through a saw-tooth like pore potential, as suggested in
\cite{lubensky} for explaining the surprisingly large translocation
times for DNA translocation. (2) In reference \cite{simon,peskin} the
chaperones bind independently along the polymer. Here, we include the
fact typically chaperones are larger than the size of a monomer,
giving rise to a ``car parking effect''. (3) We here consider
arbitrary ratios of the relevant (un)binding rate and the diffusive
rate - this is in contrast to the results in references
\cite{simon,peskin} where only the cases of Brownian ratchet motion
and fast (un)binding dynamics were considered.

An implicit assumption in our study is that the number of binding
proteins in the bath surrounding the translocating chain is large
(but of finite concentration), see reference \cite{ambme}, appendix
B. In other words, we assume that the diffusion of the chaperones in volume
is fast in comparison to the time scales of translocation and chaperone
(un)binding dynamics (well-mixing condition). Thus, chaperone bath depletion
effects, that may occur for a finite
number of chaperones \cite{zandi}, are not present in this study.
We note that we have not explicitly included chain flexibility into
the present scheme. We have also assumed that the binding energy is
the same along the polymer, and that the major friction for the
polymer originates from interactions in the pore (we hence neglected
hydrodynamic friction from the polymer part sticking out of the
pore). Although each of these effects may be relevant to experimental
situations of chaperone-driven translocation, here we focused on the
pure chaperone effect. Note that one efficient way to include
additional details in the translocation scenario is the stochastic
simulation technique based on the Gillespie algorithm
\cite{gillespie,suman}. It would also be interesting to compare the
results from this study with Brownian dynamics simulations along
similar lines as in reference \cite{zandi}, but with the
(realistically) large chaperones considered here.

We finally point out that the coupled dynamics studied here is a
generic example of how a stochastic process can rectify another
one. Our (2+1)-variable master equation formalism provides a general
framework for treating such rectification processes, that may also be
useful in other fields.

\begin{appendix}

\section{Proof of the orthogonality and the negativeness of eigenvalues}\label{proof}

In this appendix we prove an orthonormality relation for the
eigenvectors $Q_p(s)$ and that the corresponding eigenvalues $-\eta_p$ are negative, see equation \eref{eigen2}.

In terms of the running variable $s$, see equation \eref{eq:s}, and the $W$-matrix defined in equation \eref{eq:W} the detailed balance conditions
\eref{dbt} and \eref{dbr} can be written:
\begin{equation}
\label{detail}
W(s,s')P^{\mathrm{st}}(s')=W(s',s) P^{\mathrm{st}}(s),
\end{equation}
for $s\neq s'$; we note that the equation above holds
trivially also for $s=s'$. Here,
$P^{\mathrm{st}}(s)=\mathscr{Z}(s)/\mathscr{Z}$ with $\mathscr{Z}(s)$
being the partition coefficient and
$\mathscr{Z}=\sum_s\mathscr{Z}(s)$.

To derive the orthogonality relation from the detailed balance condition (\ref{detail}), consider the expression
\begin{eqnarray}
\nonumber
& &-(\eta_p-\eta_{p'})\sum_s\frac{Q_p(s)Q_{p'}(s)}{P^{\mathrm{st}}(s)}\\
&=&\sum_{s,s'}\frac{W(s,s')Q_p(s')Q_{p'}(s)}{P^{\mathrm{st}}(s)}
-\sum_{s,s'}\frac{Q_p(s)W(s,s')Q_{p'}(s')}{P^{\mathrm{st}}(s)}\nonumber\\
&=&\sum_{s,s'}\frac{W(s,s')Q_p(s')Q_{p'}(s)}{P^{\mathrm{st}}(s)}
-\sum_{s,s'}\frac{W(s',s) Q_p(s)Q_{p'}(s')}{P^{\mathrm{st}}(s')}=0.
\end{eqnarray}
where we used equation (\ref{eigen2}), and 
the detailed balance condition (\ref{detail}). Therefore, the orthogonality relation
\begin{equation}
\label{ortho}
\sum_s\frac{Q_p(s)Q_{p'}(s)}{P^{\mathrm{st}(s)}}=\delta_{p,p'},
\end{equation}
follows. Returning to the double index $(m,n)$, this is equivalent to
the statement
\begin{equation}
\sum_{m,n}\frac{Q_p(m,n)Q_{p'}(m,n)}{P^{\mathrm{st}}(m,n)}=\delta_{p,p'}.
\label{eq:ortho_relation}
\end{equation}
for the eigenfunctions $Q_p(m,n)$.

To prove that all the eigenvalues are negative it is convenient to
separate the $W$-matrix into the corresponding reflective matrix
$W_{\rm r}(s,s')$ and the additional absorbing terms
\begin{equation}
W(s,s')=W_{\rm r}(s,s')-w_{\rm a}(s)\delta_{s,s'},
\end{equation}
where the absorbing part is
\begin{equation}
w_{\rm a}(s)=-\sum_{s'}W(s',s).
\end{equation}
It follows from equation (\ref{eq:W}) that
\begin{equation}
w_{\rm a}(s)\ge 0,
\end{equation}
with at least one of the $w_{\rm a}(s)$ strictly greater than zero, and
\begin{equation}
\sum_s W_{\rm r}(s,s')=0,\label{eq:probcons}
\end{equation}
which is the condition that probability is conserved when there are no
absorbing terms.  Note that it does not matter for the detailed
balance conditions, equation (\ref{detail}), that the diagonal of the
$W$-matrix is modified. Therefore detailed balance is also satisfied
for the reflective matrix: $W_{\rm r}(s,s')P^{\mathrm{st}}(s')=W_{\rm
r}(s',s) P^{\mathrm{st}}(s)$.  This detailed balance condition
together with equation (\ref{eq:probcons}) guarantees that
$P^{\mathrm{st}}(s)$ is an eigenvector of the $W_{\rm r}$-matrix with
zero eigenvalue \cite{vankampen}. Furthermore, since the system is
ergodic, i.e. it is possible for the system to reach any state from
any other state, any eigenvector of $W_{\rm r}$ with zero eigenvalue will be
proportional to $P^{\mathrm{st}}(s)$ \cite{vankampen}.  The
negativeness of the eigenvalues $-\eta_p$ follows then because
\begin{eqnarray}
-\eta_p&=&\sum_{s,s'}\frac{W(s,s')Q_p(s')Q_p(s)}{P^{\mathrm{st}}(s)}\nonumber\\
&=&\sum_{s,s'}\frac{W_{\rm r}(s,s')Q_p(s')Q_p(s)}{P^{\mathrm{st}}(s)}-\sum_s\frac{Q_p(s)^2}{P^{\mathrm{st}}(s)}w_{\rm a}(s)<0.
\end{eqnarray}
The first term after the last equality sign is non-positive since the
$W_{\rm r}$-matrix is negative semi-definite \cite{vankampen}. The
only possibility for this term to be zero is that $Q_p(s)$ is
proportional to $P^{\mathrm{st}}(s)$. In this case, however, the last
term will be strictly negative because at least one of the $w_{\rm
a}(s)$ is non-zero.

\section{Mean first passage time}

For a one-variable master equation with reflecting boundary condition
at $m=0$ and absorbing boundary condition at $m=M+1$, and with
transfer coefficient $t^+(m)$ ($t^-(m)$) in the forward (backward)
direction, the mean first passage time is given by the explicit
expression (see reference \cite{gardiner}, chapter 7.4)
\begin{equation}
\intercal=\sum_{m=0}^M\Phi(m)\sum_{m'=0}^m\frac{1}{\mathsf{t}^+(m')\Phi(m')},\label{eq:T_1d}
\end{equation}
where
\begin{equation}
\Phi(m)=\prod_{j=1}^m\frac{\mathsf{t}^-(j)}{\mathsf{t}^+(j)}.
\end{equation}
and we have assumed that initially $m=0$. If we invoke the detailed
balance condition
$\mathsf{t}^+(m-1)\mathscr{Z}(m-1)=\mathsf{t}^-(m)\mathscr{Z}(m)$, for
$m=1,...M$, where $\mathscr{Z}(m)$ is the partition coefficient,
equation \eref{eq:T_1d} reduces to
  \be
\intercal=\sum_{m=0}^M \frac{1}{\mathsf{t}^+(m)\mathscr{Z}(m)} \sum_{m'=0}^m \mathscr{Z}(m')\label{eq:T_1d_alt}
  \ee
The expressions above are used in section \ref{sec:results}.

\end{appendix}

\end{document}